\documentclass[12pt,a4paper]{article}

\usepackage{amsmath,amssymb}
\usepackage{bm}
\usepackage{graphicx}
\usepackage{ascmac}
\usepackage{wrapfig}
\usepackage{braket}
\usepackage{color}
\usepackage{mathrsfs}
\usepackage[english]{babel}
\usepackage{inconsolata}
\usepackage{cases}
\usepackage{tikz}
\usepackage{cite}
\usepackage{setspace}
\usetikzlibrary{intersections, calc}
\usepackage[top=3cm, bottom=1.4cm, left=2.3cm, right=2.3cm, includefoot]{geometry}

\newcommand{\beq}{\begin{equation}}
\newcommand{\eeq}{\end{equation}}
\newcommand{\lr}[1]{\left(#1\right)}
\newcommand{\lrk}[1]{\left[#1\right]}
\newcommand{\Schro}{Schr$\rm{\ddot{o}}$dinger\ }

\makeatletter
\@addtoreset{equation}{section}

\makeatother

\begin{document}

\begin{titlepage}

\pagenumbering{gobble}

\begin{flushright}
\rm TIT/HEP-683 \\ March, 2021
\end{flushright}

\renewcommand{\thefootnote}{*}

\vspace{0.2in}
\begin{center}
\Large{\bf Quantum periods and TBA equations for \\ $\mathcal{N}=2\ SU(2)\ N_f=2$ SQCD with flavor symmetry}
\end{center}
\vspace{0.2in}
\begin{center}
\large Keita Imaizumi\footnote{E-mail: k.imaizumi@th.phys.titech.ac.jp}
\end{center}

\begin{center}{\it
Department of Physics,
\par
Tokyo Institute of Technology
\par
Tokyo, 152-8551, Japan
}
\end{center}
\vspace{0.2in}
\begin{abstract}
\begin{spacing}{1.0}
{\footnotesize We apply the exact WKB analysis to the quantum Seiberg-Witten curve for 4-dimensional $\mathcal{N} = 2\ SU(2)\ N_f=2$ SQCD with the flavor symmetry. The discontinuity and the asymptotic behavior of the quantum periods define a Riemann-Hilbert problem. We derive the thermodynamic Bethe ansatz (TBA) equations as a solution to this problem. We also compute the effective central charge of the underlying CFT, which is shown to be proportional to the one-loop beta function of the SQCD.}
\end{spacing}
\end{abstract}

\end{titlepage}

\newpage

\pagenumbering{arabic}
\setcounter{page}{1}

\renewcommand{\thefootnote}{\arabic{footnote}}
\setcounter{footnote}{0}

\section{Introduction}
\label{sec:intro}
The low-energy effective dynamics of 4-dimensional $\mathcal{N} = 2$ gauge theories is determined by a single holomorphic function called the prepotential \cite{Pre}. According to the Seiberg-Witten theory \cite{hep-th/9407087, hep-th/9408099}, the prepotential can be exactly computed from the Seiberg-Witten periods, which are the period integrals of the Seiberg-Witten differential on the Seiberg-Witten curve describing the Coulomb moduli space of the vacua. The prepotential obtained from the Seiberg-Witten periods enables us to understand non-perturbative aspects of the gauge theories such as the global structure of the BPS spectra \cite{hep-th/9602082, hep-th/9605101, hep-th/9611012, hep-th/9706145}. The prepotential can also be obtained from the Nekrasov partition function \cite{hep-th/0206161, hep-th/0306238}, which is defined on the $\Omega$-deformed background parametrized by two deformation parameters $\epsilon$ and $\epsilon'$. 
\par
Under the $\Omega$-background, the prepotential and the Seiberg-Witten periods also receive the correction by the deformation. In \cite{0910.5670}, it was argued that the deformed Seiberg-Witten periods for the $SU(2)$ super Yang-Mills theory in the Nekrasov-Shatashvili limit $\epsilon' \rightarrow 0$ \cite{0908.4052} are identical to the quantum periods (or the WKB periods) for the Mathieu differential equation. This identification was generalized to pure $SU(N)$ case \cite{0911.2396} and with matters \cite{1103.4843} later. The corresponding differential equations are called the quantum Seiberg-Witten curves, which are obtained from the canonical quantization of the symplectic structure of the Seiberg-Witten curves. The quantization of the Seiberg-Witten curves has been investigated with various examples \cite{1705.09120, 1804.04815, 1903.00168, 2001.08891}. The quantum Seiberg-Witten curves also appear in the AGT correspondence, where the differential equations are satisfied by the one-point function of a degenerate primary field with respect to the Gaiotto states \cite{0910.4431, 1008.0574}.
\par
Recently, there have been remarkable developments for the resurgent perspective of the quantum Seiberg-Witten curves. In \cite{1504.08324}, the authors applied the exact WKB analysis to the quantum Seiberg-Witten curve for the pure $SU(2)$ theory and determined the instanton corrections to the prepotential. The similar analysis for 4d $\mathcal{N}=2$ $SU(2)$ SQCD with $N_f \leq 4$ fundamental hypermultiplets is presented in \cite{1604.05520}, which is based on the quantum Seiberg-Witten curves obtained from the AGT correspondence.
\par
The exact WKB analysis for the quantum Seiberg-Witten curves also has a relation to the 2-dimensional integrable QFT. In \cite{1811.04812}, the authors investigated the exact WKB analysis for the quantum Seiberg-Witten curves of the $(A_1, A_r)$ Argyres-Douglas theory (e.g.\cite{1803.02320, 1806.01407}) and derived the thermodynamic Bethe ansatz (TBA) equations governing the Borel resummations of the quantum periods as the solution to a Riemann-Hilbert problem explained by Voros \cite{Vor}. The TBA equations for the quantum periods are also derived for the pure $SU(2)$ theory \cite{1908.07065, 2002.06829} and the $(A_1, D_r)$ Argyres-Douglas theory \cite{1910.09406}. The exact WKB analysis and the TBA equations for the quantum Seiberg-Witten curves are also studied in the context of the abelianization (e.g. \cite{1906.04271, 2012.15658}). 
\par
It is important to generalize the $SU(2)$ SYM to the SQCD, which has the moduli space with higher dimensions. The quantum Seiberg-Witten curves for 4d $\mathcal{N}=2$ $SU(2)$ SQCD with $N_f \leq 4$ can be obtained from the quantization of the Seiberg-Witten curves \cite{1705.09120}. The quantum curves are also obtained from the AGT correspondence with the mass decoupling limits, which decouple the different hypermultiplets considered in \cite{1604.05520}. In this paper, we derive the TBA equations for $N_f=2$ SQCD with the flavor symmetry based on the curve in \cite{1705.09120}. The moduli space of this gauge theory has three important limits, the massless limit, the decoupling limit and the Argyres-Douglas limit. For the massless limit and the decoupling limit, the quantum SW curve becomes the one for the pure $SU(2)$ case and the TBA equations for this curve are already derived in \cite{1908.07065, 2002.06829}. The TBA equations for the Argyres-Douglas limit are also derived in \cite{1811.04812}. Therefore, by deriving the TBA equations for $N_f=2$ SQCD with the flavor symmetry, we can study these flow of the theory from the point of view of the TBA equations.
\par
This paper is organized as follows. In section \ref{sec:EWKB}, we apply the exact WKB analysis to the quantum Seiberg-Witten curve and define a Riemann-Hilbert problem. In section \ref{sec:TBA}, we derive the TBA equations as a solution to the Riemann-Hilbert problem. We then study some special limits of the TBA equations, the massless limit, the decoupling limit and the Argyres-Douglas limit. We also compute the effective central charge of the underlying CFT, which is found to be proportional to the coefficient of the one-loop beta function of the SQCD.

\section{Exact WKB analysis and Quantum SW curve}
\label{sec:EWKB}
The quantum Seiberg-Witten curve for 4-dimensional $\mathcal{N} = 2\ SU(2)$ SQCD with two fundamental hypermultiplets take the form of the \Schro type differential equation \cite{1705.09120}, 
\beq
\label{eq:QSWC}
 \lr{-\epsilon^2\frac{d^2}{dq^2} + \lr{V\lr{q} - u}}\psi\lr{q} = 0, \ \ \ V\lr{q} = -\frac{\Lambda_2}{2}\lr{m_1e^{iq} + m_2e^{-iq}} - \frac{\Lambda_2^2}{8}\cos\lr{2q},
\eeq
where $u$ is the Coulomb moduli parameter, $\Lambda_2$ is the dynamically generated scale, $m_1, m_2$ are the bare masses of the hypermultiplets, $\epsilon$ is the deformation parameter in the Nekrasov-Shatashvili limit of the $\Omega$-background and $q$ is a complex variable. In the context of the AGT-correspondence, the variable $q$ is the position of a surface operator in the 4d gauge theory (see e.g. \cite{1008.0574}).
\par
The standard WKB method produces an asymptotic expansion in $\epsilon$ of the solution to (\ref{eq:QSWC}),
\beq
\label{eq:evensol}
 \psi(q) = \frac{1}{\sqrt{P_{\mathrm{even}}(q)}}\exp(\frac{i}{\epsilon}\int^{q}P_{\mathrm{even}}(q')dq'), 
\eeq
where
\beq
\label{eq:evod}
 P_{\mathrm{even}}(q) = \sum_{n=0}^{\infty}p_{2n}(q)\epsilon^{2n}.
\eeq
By substituting the solution (\ref{eq:evensol}) into (\ref{eq:QSWC}), we can obtain $p_{2n}(q)$ recursively. Especially we find $p_0(q) = \sqrt{u - V(q)}$.
\par
$P_{\mathrm{even}}(q)dq$ in (\ref{eq:evensol}) can be regarded as a one-form on the Riemann surface $\Sigma$ defined by the following algebraic curve,
\begin{equation}
 y^2 = u - V(q).
\end{equation}
We will call $\Sigma$ WKB curve. The Riemann surface $\Sigma$ is the same as the one defined by the Seiberg-Witten curve of the $SU(2)\ N_f=2$ theory \cite{1705.09120}. The one-cycles $\gamma \in H_1(\Sigma)$ generate the periods of $P_{\mathrm{even}}(q)dq$,
\begin{equation}
 \Pi_{\gamma} := \oint_{\gamma} P_{\mathrm{even}}(q)dq, \ \ \ \ \ \ \gamma \in H_1(\Sigma),
\end{equation}
which we will call quantum periods. The quantum periods determine the low-energy effective dynamics of the SQCD. As $P_{\mathrm{even}}(q)$ goes, the quantum periods are even power series in $\epsilon$,
\begin{equation}
 \Pi_{\gamma} = \sum_{n=0}^{\infty} \Pi_{\gamma}^{(n)}\epsilon^{2n},\ \ \ \ \ \ \Pi_{\gamma}^{(n)} = \oint_{\gamma}p_{2n}(q)dq. 
\end{equation}
We can compute the higher order corrections to the quantum periods by using the differential operator technique \cite{1705.09120}:
\beq
\label{eq:difop}
 \Pi_{\gamma}^{(n)} = \mathcal{O}_{n}\Pi_{\gamma}^{(0)},
\eeq
where $\mathcal{O}_{n}$ is a differential operator with respect to the moduli parameters on the WKB curve.
\par
The quantum periods are asymptotic series, which converge only at $|\epsilon|=0$, and therefore need to be properly resumed. In the exact WKB analysis, we take Borel resummation technique. First, we define Borel transformation of a quantum period as follows:
\begin{equation}
 \hat{\Pi}_{\gamma}(\xi) = \sum_{n=0}^{\infty}\frac{1}{(2n)!}\Pi_{\gamma}^{(n)}\xi^{2n},
\end{equation}
where $\xi$ is a complex variable. With the help of the factor $1/(2n)!$, the Borel transformation has a finite convergence of radius. Therefore the Borel transformation can be analytically continued on the whole of $\xi$-plane. The Borel resummation of the quantum period is  then defined by the Laplace integral of the Borel transformation,
\begin{equation}
\label{eq:Bsum}
 s_{\varphi}\left(\Pi_{\gamma}\right)\left(\epsilon\right) = \frac{1}{\epsilon}\int_{0}^{e^{i\varphi}\infty} e^{-\xi/\epsilon}\hat{\Pi}_{\gamma}(\xi)d\xi,
\end{equation}
where $\varphi$ is the phase of $\epsilon$ ($\epsilon = |\epsilon|e^{i\varphi}$). The Borel resummation of the quantum period $s_{\varphi}\left(\Pi_{\gamma}\right)\left(\epsilon\right)$ is an analytic function and has the quantum period $\Pi_{\gamma}$ as the asymptotic expansion in $\epsilon \rightarrow 0$.
\par
In general, the analytic continuations of the Borel transformations have singularity, which are typically poles and blanch cuts, on the $\xi$-plane. If the Borel transformation of a quantum period $\Pi_{\gamma}$ has singularities on the ray along a direction in the $\xi$-plane, then the integral (\ref{eq:Bsum}) cannot be defined in this direction. In that case, instead of (\ref{eq:Bsum}), we use the integrals which avoid the singularities to the left or right,
\begin{equation}
\label{eq:lBsum}
 s_{\varphi\pm}\left(\Pi_{\gamma}\right)\left(e^{i\varphi}|\epsilon|\right) = \lim_{\delta \rightarrow +0} s_{\varphi \pm \delta}\left(\Pi_{\gamma}\right)\left(e^{i(\varphi \pm \delta)}|\epsilon|\right).
\end{equation}
The discontinuity of a quantum period $\Pi_{\gamma}$ is then given as the difference of (\ref{eq:lBsum}),
\begin{equation}
 \mathrm{disc}_{\varphi} \Pi_{\gamma} := s_{\varphi+}\left(\Pi_{\gamma}\right) - s_{\varphi-}\left(\Pi_{\gamma}\right). 
\end{equation}

\begin{figure}[tp]
  \begin{center}
  \hspace*{0.3in}
   \begin{tikzpicture}
    \draw[->,>=stealth,semithick] (-6,0)--(6,0) node[right]{\Large$\mathrm{Re}\lr{q}$};
    \draw[->,>=stealth,semithick] (0,-3)--(0,3) node[above]{\Large$V\lr{q}$};
    \draw[thick] (-5.5,0.2)--(-5.5,-0.2);
    \draw (-5.5,-0.2) node[below]{\large$-\pi$};
    \draw[thick] (5.5,0.2)--(5.5,-0.2);
    \draw (5.5,-0.2) node[below]{\large$\pi$};
    \draw[thick] (-2.75,0.2)--(-2.75,-0.2);
    \draw (-2.75,-0.2) node[below]{\large$-\frac{\pi}{2}$};
    \draw[thick] (2.75,0.2)--(2.75,-0.2);
    \draw (2.75,-0.2) node[below]{\large$\frac{\pi}{2}$};
    \draw (0,0)node[below right]{\large$0$};
    \draw[samples=100,domain=-6:6,line width=1pt] plot(\x,{-(1/2)*cos(pi*\x/(5.5) r)-2*cos(2*pi*\x/(5.5) r)});
    \draw[red, thick] (-5.5,1)--(5.5,1) node[right]{\Large$u$};
    \draw[dashed, thick] (-3.8,1)--(-3.8,0);
    \draw (-3.8,0) node[below]{\large$-q_2$};
    \draw[dashed, thick] (3.8,1)--(3.8,0);
    \draw (3.8,0) node[below]{\large$q_2$};
    \draw[dashed, thick] (1.95,1)--(1.95,0);
    \draw (1.95,0) node[below]{\large$q_1$};
    \draw[dashed, thick] (-1.95,1)--(-1.95,0);
    \draw (-1.95,0) node[below]{\large$-q_1$};
   \end{tikzpicture}
   \caption{The potential on the real axis ($\Lambda_2 = 4$, $m = 1/8$ and $u = 1$). $\lr{q_1, q_2, -q_1, -q_2}$ are the solutions to $u = V(q_*)$.}
   \label{fig:pot}
  \end{center}
\end{figure}
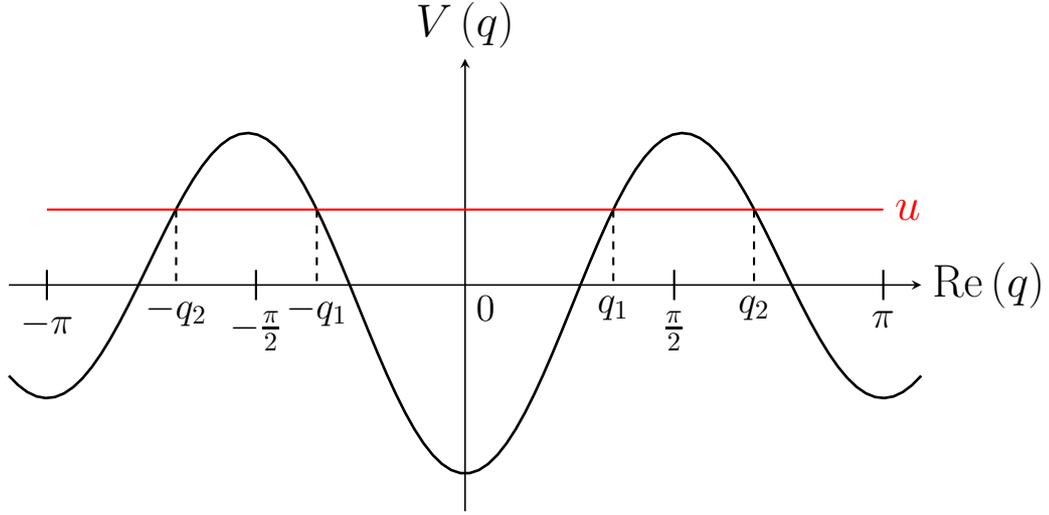

Let us compute the discontinuity of the quantum periods for (\ref{eq:QSWC}). We take a special region on the moduli space in which all the solutions to $u = V(q_*)$ become real and different. This region can be realized as follows: We consider the case that  the hypermultiplets have a same mass $m_1 = m_2 = m$ and $m, u, \Lambda_2 \in \mathbb{R}$. Then $u - V(q)$ becomes a real function on the $\Re(q)$ axis,
\beq
\label{eq:samemass}
 u - V\lr{q} = u + \Lambda_2m\cos\lr{q} + \frac{\Lambda_2^2}{8}\cos\lr{2q}.
\eeq
Restricting the parameters to $-\frac{\Lambda_2}{2} < m < \frac{\Lambda_2}{2}$ and $\Lambda_2|m| - \frac{\Lambda_2^2}{8} < u < m^2 + \frac{\Lambda_2^2}{8}$, the solutions to $u = V(q_*)$ become real and different (see Fig.\ref{fig:pot})\footnote{This configuration corresponds to the bound state in the context of quantum mechanics.}.
In the context of the gauge theory, the restriction of $u$ corresponds to the strong coupling region.
\par
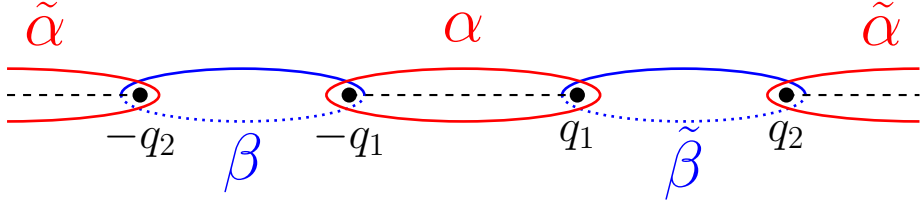
\begin{figure}[tp]
  \begin{center}
   \begin{tikzpicture}
    \fill (-4.25,0) circle [radius=0.1];
    \draw (-4.25,-0.2) node[below]{\Large$-q_2$};
    \fill (4.25,0) circle [radius=0.1];
    \draw (4.25,-0.2) node[below]{\Large$q_2$};
    \fill (-1.5,0) circle [radius=0.1];
    \draw (-1.5,-0.2) node[below]{\Large$-q_1$};
    \fill (1.5,0) circle [radius=0.1];
    \draw (1.5,-0.2) node[below]{\Large$q_1$};
    \draw[blue, line width=1pt] (-1.3,0) arc (0:180: 1.6cm and 0.35cm);
    \draw[dotted, blue, line width=1pt] (-4.5,0) arc (180:360: 1.6cm and 0.35cm)node at (-2.9,-0.9) {\huge$\beta$};
    \draw[blue, line width=1pt] (4.5,0) arc (0:180: 1.6cm and 0.35cm);
    \draw[dotted, blue, line width=1pt] (1.3,0) arc (180:360: 1.6cm and 0.35cm)node at (2.9,-0.9) {\huge$\tilde{\beta}$};
    \draw[red, line width=1pt] (0,0) circle [x radius=1.8cm, y radius=3.5mm] node at (0,0.9) {\huge$\alpha$};
    \draw[red, line width=1pt] (-4.0,0) arc (0:90: 2.0cm and 0.35cm);
    \draw[red, line width=1pt] (-6.0,-0.35) arc (270:360: 2.0cm and 0.35cm)node at (-5.5,0.95) {\huge$\tilde{\alpha}$};
    \draw[red, line width=1pt] (4.0,0) arc (180:90: 2.0cm and 0.35cm);
    \draw[red, line width=1pt] (4.0,0) arc (180:270: 2.0cm and 0.35cm)node at (5.5,0.95) {\huge$\tilde{\alpha}$};
    \draw[dashed, thick] (-6,0)--(-4.25,0);
    \draw[dashed, thick] (-1.5,0)--(1.5,0);
    \draw[dashed, thick] (4.25,0)--(6,0);
   \end{tikzpicture}
   \caption{The one-cycles on a sheet of the WKB curve. The dashed black lines indicate branch cuts.}
   \label{fig:cycles}
  \end{center}
\end{figure}

One can then finds four independent cycles on the WKB curve, $\alpha$, $\tilde{\alpha}$ cycles, encircling the classically allowed intervals, and $\beta$, $\tilde{\beta}$ cycles, encircling the classically forbidden intervals (Fig.\ref{fig:cycles}). We choose the orientations of the cycles so that 
\begin{equation}
\label{eq:clper}
\begin{split}
 & Z_1 := \Pi^{(0)}_{\alpha} = \oint_{\alpha} p_0(q)dq, \ \ \ Z_2 := i\Pi_{\beta}^{(0)} = i\oint_{\beta}p_0(q)dq, \\
 & Z_3 := \Pi^{(0)}_{\tilde{\alpha}} = \oint_{\tilde{\alpha}} p_0(q)dq, \ \ \ Z_4 := i\Pi_{\tilde{\beta}}^{(0)} = i\oint_{\tilde{\beta}}p_0(q)dq,
\end{split}
\end{equation}
are real and positive.
\par
Under the change of variable $e^{iq} = z$, $p_0(q)dq$ becomes the root of the quadratic differential for 4d $N=2\ SU(2)\ N_f = 2$ gauge theory considered in \cite{0907.3987}. The quadratic differential has two irregular singularities at $z = 0, \infty$, which correspond to $q = \pm i\infty$. At these points, $2\pi i\mathrm{Res}(\pm i\infty, p_0) = 2\pi m$ and $\Pi_{\alpha - \tilde{\alpha}}^{(0)}$ becomes the sum of these residues, while $\Pi_{\beta - \tilde{\beta}}^{(0)}$ becomes the difference of these residues (namely zero). Therefore we obtain the following relation,
\beq
\label{eq:idn}
 \Pi_{\alpha}^{(0)} = \Pi_{\tilde{\alpha}}^{(0)} + 4\pi m, \ \ \ \Pi_{\beta}^{(0)} = \Pi_{\tilde{\beta}}^{(0)}.
\eeq
Then one can show that the higher order coefficients are equivalent respectively,
\beq
\label{eq:idhigh}
 \Pi_{\alpha}^{(n)} = \Pi_{\tilde{\alpha}}^{(n)}, \ \ \ \Pi_{\beta}^{(n)} = \Pi_{\tilde{\beta}}^{(n)},\ \ \ (n \geq 1)
\eeq
because the differential operator $\mathcal{O}_n$ can be expressed by only using $u$-derivative \cite{1705.09120}.

The discontinuity of the Borel resummations of the quantum periods can be captured by the Delabaere-Dillinger-Pham formula (theorem 2.5.1 of \cite{DP}, and theorem 3.4 of \cite{cr1}). For $\varphi = 0$, the formula says that the periods for the classically allowed intervals have the discontinuity, and it is encoded in the periods for the classically forbidden intervals,
\begin{equation}
\label{eq:DDP}
 \frac{i}{\epsilon}\mathrm{disc}_{0}\ \Pi_{\gamma_{\mathrm{ca}}} = -\sum_{\gamma_{\mathrm{cf}}}\langle\gamma_{\mathrm{ca}} , \gamma_{\mathrm{cf}}\rangle\log\left[1 + \exp\left(-\frac{i}{\epsilon}s_0(\Pi_{\gamma_{\mathrm{cf}}})\right)\right],
\end{equation}
where $\gamma_{\mathrm{ca}}$ is a classically allowed cycle, $\gamma_{\mathrm{cf}}$ is the classically forbidden cycles and $\langle\gamma_{\mathrm{ca}} , \gamma_{\mathrm{cf}}\rangle$ is the intersection number of them. By using this formula, the discontinuities of $\Pi_{\alpha}$, $\Pi_{\tilde{\alpha}}$ are given as follows,
\beq
\label{eq:a0disc}
 \frac{i}{\epsilon}\mathrm{disc}_{0}\ \Pi_{\alpha} = -\log\left[1 + \exp\left(-\frac{i}{\epsilon}s_0(\Pi_{\beta})\right)\right] -\log\left[1 + \exp\left(-\frac{i}{\epsilon}s_0(\Pi_{\tilde{\beta}})\right)\right],
\eeq
\beq
\label{eq:at0disc}
 \frac{i}{\epsilon}\mathrm{disc}_{0}\ \Pi_{\tilde{\alpha}} = -\log\left[1 + \exp\left(-\frac{i}{\epsilon}s_0(\Pi_{\beta})\right)\right] -\log\left[1 + \exp\left(-\frac{i}{\epsilon}s_0(\Pi_{\tilde{\beta}})\right)\right].
\eeq
\par
$\Pi_{\alpha}$ and $\Pi_{\tilde{\alpha}}$ also have the discontinuity for the direction $\varphi = \pi$ because the quantum periods are even power series in $\epsilon$. Similarly, $\Pi_{\beta}$ and $\Pi_{\tilde{\beta}}$ have the discontinuity for the direction $\varphi = \pm \frac{\pi}{2}$ because, in this direction, the classically allowed intervals and classically forbidden intervals are switched.
\par
The asymptotic behavior and the discontinuities for the Borel resummations of the quantum periods define a Riemann-Hilbert problem for themselves \cite{Vor}. In the next section, we derive the TBA equations as a solution to this problem.

\section{Quantum periods and TBA equations}
\label{sec:TBA}

\subsection{TBA equations}
\label{sec:DerTBA}
The discontinuities obtained in the previous section can be put into a uniform description by introducing functions $\varepsilon_i\lr{\theta}$ as
\begin{equation}
\label{eq:pseudo}
\begin{split}
 & \varepsilon_{1}\left(\theta + \frac{i\pi}{2} - i\varphi\right) = \frac{i}{\epsilon}s_{\varphi}\left(\Pi_{\alpha}\right)\left(\epsilon\right),\ \ \ \ \  \varepsilon_{2}\left(\theta - i\varphi \right) = \frac{i}{\epsilon}s_{\varphi}\left(\Pi_{\beta}\right)\left(\epsilon\right), \\
 & \varepsilon_{3}\left(\theta + \frac{i\pi}{2} - i\varphi\right) = \frac{i}{\epsilon}s_{\varphi}\left(\Pi_{\tilde{\alpha}}\right)\left(\epsilon\right),\ \ \ \ \  \varepsilon_{4}\left(\theta - i\varphi \right) = \frac{i}{\epsilon}s_{\varphi}\left(\Pi_{\tilde{\beta}}\right)\left(\epsilon\right),
\end{split}
\end{equation}
where $\theta$ is defined by
\begin{equation}
 \frac{1}{\epsilon} = e^{\theta - i\varphi}.
\end{equation}
The discontinuities are then put together into a simpler one,
\begin{equation}
\label{eq:discep}
 \mathrm{disc}_{\pm\pi/2}\varepsilon_i\left(\theta\right) = \pm\left[L_{i-1}(\theta) + L_{i+1}(\theta)\right], \ \ \ \ (i = 1, 2, 3, 4)
\end{equation}
where
\begin{equation}
 L_{i}(\theta) = \log\left(1 + e^{-\varepsilon_i(\theta)}\right)
\end{equation}
and we define $L_0 = L_4$, $L_5 = L_1$.
\par
The functions $\varepsilon_i\lr{\theta}$ have the following asymptotic behavior,
\begin{equation}
\label{eq:asyep}
  \varepsilon_i(\theta) = Z_ie^{\theta} + \mathcal{O}(e^{-\theta}), \ \ \ \theta \rightarrow \infty, 
\end{equation}
because $\theta \rightarrow \infty$ expansion of $\varepsilon_i\lr{\theta}$ is equivalent to $\epsilon \rightarrow 0$ expansion of the Borel resummations of the quantum periods.
\par
Now we can derive the TBA equations for the functions $\varepsilon_i\lr{\theta}$ satisfying the conditions (\ref{eq:discep}) and (\ref{eq:asyep}) \cite{1811.04812}. For the present case, the solution is given by the following TBA system,
\begin{equation}
\label{eq:TBA}
\begin{split}
 \varepsilon_1(\theta) &= Z_1e^{\theta} -\int_{\mathbb{R}}\frac{\log\left(1 + e^{-\varepsilon_2(\theta')}\right)}{\cosh (\theta - \theta')}\frac{d\theta'}{2\pi} -\int_{\mathbb{R}}\frac{\log\left(1 + e^{-\varepsilon_4(\theta')}\right)}{\cosh (\theta - \theta')}\frac{d\theta'}{2\pi}, \\
 \varepsilon_2(\theta) &= Z_2e^{\theta} -\int_{\mathbb{R}}\frac{\log\left(1 + e^{-\varepsilon_1(\theta')}\right)}{\cosh (\theta - \theta')}\frac{d\theta'}{2\pi} -\int_{\mathbb{R}}\frac{\log\left(1 + e^{-\varepsilon_3(\theta')}\right)}{\cosh (\theta - \theta')}\frac{d\theta'}{2\pi}, \\
 \varepsilon_3(\theta) &= Z_3e^{\theta} -\int_{\mathbb{R}}\frac{\log\left(1 + e^{-\varepsilon_2(\theta')}\right)}{\cosh (\theta - \theta')}\frac{d\theta'}{2\pi} -\int_{\mathbb{R}}\frac{\log\left(1 + e^{-\varepsilon_4(\theta')}\right)}{\cosh (\theta - \theta')}\frac{d\theta'}{2\pi}, \\
 \varepsilon_4(\theta) &= Z_4e^{\theta} -\int_{\mathbb{R}}\frac{\log\left(1 + e^{-\varepsilon_1(\theta')}\right)}{\cosh (\theta - \theta')}\frac{d\theta'}{2\pi} -\int_{\mathbb{R}}\frac{\log\left(1 + e^{-\varepsilon_3(\theta')}\right)}{\cosh (\theta - \theta')}\frac{d\theta'}{2\pi}.
\end{split}
\end{equation}
The identification (\ref{eq:idn}) leads to $\varepsilon_2(\theta) = \varepsilon_4(\theta)$ and the TBA equations can be collapsed to three,
\begin{equation}
\label{eq:TBAid}
\begin{split}
 \varepsilon_1(\theta) &= Z_1e^{\theta} -2\int_{\mathbb{R}}\frac{\log\left(1 + e^{-\varepsilon_2(\theta')}\right)}{\cosh (\theta - \theta')}\frac{d\theta'}{2\pi}, \\
 \varepsilon_2(\theta) &= Z_2e^{\theta} -\int_{\mathbb{R}}\frac{\log\left(1 + e^{-\varepsilon_1(\theta')}\right)}{\cosh (\theta - \theta')}\frac{d\theta'}{2\pi} -\int_{\mathbb{R}}\frac{\log\left(1 + e^{-\varepsilon_3(\theta')}\right)}{\cosh (\theta - \theta')}\frac{d\theta'}{2\pi}, \\
 \varepsilon_3(\theta) &= Z_3e^{\theta} -2\int_{\mathbb{R}}\frac{\log\left(1 + e^{-\varepsilon_2(\theta')}\right)}{\cosh (\theta - \theta')}\frac{d\theta'}{2\pi}.
\end{split}
\end{equation}
\par
So far we only consider the special parameter region, but we can also derive the TBA equations for a pure imaginary region ($m, \Lambda_2 \in i\mathbb{R}$) and an anti-same mass region ($m_1 = -m_2$) in the same way. For general region, we need to analytically continue the TBA equations \cite{1002.2459}. Let us consider the analytic continuation for (\ref{eq:TBA}). Taking $u, m, \Lambda_2$ as complex values, $Z_i$ also becomes complex,
\begin{equation}
 Z_i = |Z_i|e^{i\phi_i}, \ \ (i = 1 \sim 4).
\end{equation}
Then we obtain the following TBA equations, 
\begin{equation}
\label{eq:cTBA}
\begin{split}
 \tilde{\varepsilon}_1(\theta) &= |Z_1|e^{\theta} -\int_{\mathbb{R}}\frac{\log\left(1 + e^{-\tilde{\varepsilon}_2(\theta')}\right)}{\cosh (\theta - \theta' + i\phi_2 - i\phi_1)}\frac{d\theta'}{2\pi} -\int_{\mathbb{R}}\frac{\log\left(1 + e^{-\tilde{\varepsilon}_4(\theta')}\right)}{\cosh (\theta - \theta' + i\phi_4 - i\phi_1)}\frac{d\theta'}{2\pi}, \\
 \tilde{\varepsilon}_2(\theta) &= |Z_2|e^{\theta} -\int_{\mathbb{R}}\frac{\log\left(1 + e^{-\tilde{\varepsilon}_1(\theta')}\right)}{\cosh (\theta - \theta' + i\phi_1 - i\phi_2)}\frac{d\theta'}{2\pi} -\int_{\mathbb{R}}\frac{\log\left(1 + e^{-\tilde{\varepsilon}_3(\theta')}\right)}{\cosh (\theta - \theta' + i\phi_3 - i\phi_2)}\frac{d\theta'}{2\pi}, \\
 \tilde{\varepsilon}_3(\theta) &= |Z_3|e^{\theta} -\int_{\mathbb{R}}\frac{\log\left(1 + e^{-\tilde{\varepsilon}_2(\theta')}\right)}{\cosh (\theta - \theta' + i\phi_2 - i\phi_3)}\frac{d\theta'}{2\pi} -\int_{\mathbb{R}}\frac{\log\left(1 + e^{-\tilde{\varepsilon}_4(\theta')}\right)}{\cosh (\theta - \theta' + i\phi_4 - i\phi_3)}\frac{d\theta'}{2\pi}, \\
 \tilde{\varepsilon}_4(\theta) &= |Z_4|e^{\theta} -\int_{\mathbb{R}}\frac{\log\left(1 + e^{-\tilde{\varepsilon}_1(\theta')}\right)}{\cosh (\theta - \theta' + i\phi_1 - i\phi_4)}\frac{d\theta'}{2\pi} -\int_{\mathbb{R}}\frac{\log\left(1 + e^{-\tilde{\varepsilon}_3(\theta')}\right)}{\cosh (\theta - \theta' + i\phi_3 - i\phi_4)}\frac{d\theta'}{2\pi},
\end{split}
\end{equation}
where 
\begin{equation}
 \tilde{\varepsilon_i}(\theta) := \varepsilon_i(\theta -i\phi_i)\ \ \ (i = 1 \sim 4).
\end{equation}
The TBA equations (\ref{eq:cTBA}) are only valid for the region $|\phi_i - \phi_j| < \frac{\pi}{2}$ because the integrands have the pole at $|\phi_i - \phi_j| = \frac{\pi}{2}$. 
For $|\phi_i - \phi_j| > \frac{\pi}{2}$, the residue of the pole deforms the TBA equations and typically we find an infinite number of the integral equations. In the language of the exact WKB analysis, the appearance of an infinite number of the TBA equations corresponds to the appearance of an infinite number of the quantum periods.
\par
The number of the quantum periods is equivalent to the number of the stable BPS states in the gauge theory \cite{0807.4723,0907.3987}. For the 4d $N=2\ SU(2)\ N_f = 2$ gauge theory, there are four stable BPS states in the strong coupling region and an infinite number of the states in the weak coupling region. Therefore we conclude that the TBA equations (\ref{eq:cTBA}) are valid for the strong coupling region.
\par
In principle, we can also derive the TBA equations for $m_1 \neq m_2$ by solving the Riemann-Hilbert problem. But in this case the potential $V(q)$ is not a real-valued function on the $\Re(q)$ axis and there is  no way to analytically determine the directions of the discontinuity as the same-mass case. This difficulty also exists for 4d $N=2\ SU(2)\ N_f =1, 3, 4$ gauge theories. However, the number of the stable BPS states for the 4d $N=2\ SU(2)\ N_f = 2$ gauge theory with $m_1 \neq m_2$ is the same for the same-mass case. Therefore we conjecture that the TBA equations (\ref{eq:cTBA}) are also valid for the $m_1 \neq m_2$ case.
\par
Now we discuss the TBA equations at some special points in the moduli space. In the massless case $m = 0$, (\ref{eq:idn}) leads to $\Pi_{\alpha} = \Pi_{\tilde{\alpha}}$ or equivalently $\varepsilon_1(\theta) = \varepsilon_3(\theta)$ and the TBA equations can be collapsed to two,
\begin{equation}
\label{eq:mslsTBA}
\begin{split}
 \varepsilon_1(\theta) &= Z_1e^{\theta} -2\int_{\mathbb{R}}\frac{\log\left(1 + e^{-\varepsilon_2(\theta')}\right)}{\cosh (\theta - \theta')}\frac{d\theta'}{2\pi}, \\
 \varepsilon_2(\theta) &= Z_2e^{\theta} -2\int_{\mathbb{R}}\frac{\log\left(1 + e^{-\varepsilon_1(\theta')}\right)}{\cosh (\theta - \theta')}\frac{d\theta'}{2\pi}.
\end{split}
\end{equation}
(\ref{eq:mslsTBA}) agrees with the TBA equations for the Mathieu equation \cite{1908.07065, 2002.06829}. This agreement is compatible with that the quantum SW curve (\ref{eq:QSWC}) becomes the Mathieu equation in the massless case. We also obtain the TBA equations (\ref{eq:mslsTBA}) in the decoupling limit ($m \rightarrow \infty$ and $\Lambda_2 \rightarrow 0$ while $m\Lambda_2$ being fixes), which turns the $N_f = 2$ theory into the pure gauge theory, because in this limit the quantum SW curve (\ref{eq:QSWC}) becomes the one for the pure $SU(2)$ theory.
\par
The point $m = \frac{\Lambda_2}{2}$, $u = \frac{3}{8}\Lambda^2$ is the superconformal or Argyres-Douglas point where mutually nonlocal BPS states become massless \cite{hep-th/9511154}. In the limit $m \rightarrow \frac{\Lambda_2}{2}$, $u \rightarrow \frac{3}{8}\Lambda^2 $ with keeping the theory in the strong coupling region, $\tilde{\alpha}$-cycle shrinks and $\Pi_{\tilde{\alpha}}$ goes to zero. If we neglect $\tilde{\alpha}$-cycle, we obtain the following TBA equations,
\begin{equation}
\label{eq:TBAD}
\begin{split}
 \varepsilon_1(\theta) &= Z_1e^{\theta} -2\int_{\mathbb{R}}\frac{\log\left(1 + e^{-\varepsilon_2(\theta')}\right)}{\cosh (\theta - \theta')}\frac{d\theta'}{2\pi}, \\
 \varepsilon_2(\theta) &= Z_2e^{\theta} -\int_{\mathbb{R}}\frac{\log\left(1 + e^{-\varepsilon_1(\theta')}\right)}{\cosh (\theta - \theta')}\frac{d\theta'}{2\pi}.
\end{split}
\end{equation}
This TBA system agrees with the TBA equations for the quartic potential derived in \cite{1811.04812}, which is the quantum SW curve for $(A_1, A_3)$ AD theory. Moreover, under the universality of the AD theory \cite{1903.00168}, the quantum SW curve for $(A_1, A_3)$ AD theory is equivalent to the $SU(2)\ N_f=2$ AD theory.
\vspace{0.2in}
\par
The large $\theta$ expansion of the TBA equations provides the all-older asymptotic expansion of the epsilon functions. For example, for (\ref{eq:TBA}),
\begin{equation}
\label{eq:asym}
  \varepsilon_i(\theta) = Z_ie^{\theta} + \sum_{n=1}^{\infty}Z_i^{(n)}e^{(1-2n)\theta}, \ \ \ \theta \rightarrow \infty,
\end{equation}
where
\begin{equation}
\label{eq:allorder}
 Z_k^{(n)} = \frac{(-1)^n}{\pi}\int_{\mathbb{R}} e^{(2n-1)\theta}(L_{k-1}(\theta) + L_{k+1}(\theta))d\theta,  \ \ \ (k = 1, 2, 3, 4).
\end{equation}
$Z_k^{(n)}$ can be replaced with the coefficients of the quantum periods as follows:
\begin{equation}
\label{eq:Eper}
 Z_{1}^{(n)} = (-1)^n\Pi_{\alpha}^{(n)}, \ \ \ \ \ Z_{2}^{(n)} = i\Pi_{\beta}^{(n)}, \ \ \ \ \ Z_{3}^{(n)} = (-1)^n\Pi_{\tilde{\alpha}}^{(n)}, \ \ \ \ \ Z_{4}^{(n)} = i\Pi_{\tilde{\beta}}^{(n)}.
\end{equation}
Moreover, (\ref{eq:allorder}) indicates the following identifications:
\beq
\label{eq:hid}
 Z_{1}^{(n)} = Z_{3}^{(n)}, \ \ \ \ \ Z_{2}^{(n)} = Z_{4}^{(n)} .
\eeq
These agree with (\ref{eq:idhigh}).
\par
We compare the calculation of the quantum periods by using the TBA equations and the differential operator (\ref{eq:difop}). In the same mass case, the first and second orders can be calculated by using the following differential operators \cite{1705.09120}:
\beq
\label{eq:FOOP}
 \mathcal{O}_{1} = \frac{1}{6}\frac{\partial}{\partial u} + \frac{1}{3}u\frac{\partial^2}{\partial u^2} + \frac{1}{4}m\frac{\partial^2}{\partial m\partial u},
\eeq 
\beq
 \mathcal{O}_{2} = \frac{5}{24}\frac{\partial^2}{\partial u^2} + \frac{1}{3}u\frac{\partial^3}{\partial u^3} + \frac{7}{90}u\frac{\partial^4}{\partial u^4} + \frac{7}{60}um\frac{\partial^4}{\partial m \partial u^3} +\frac{23}{96}m\frac{\partial^3}{\partial m^2 \partial u^2} + \frac{7}{160}m^2\frac{\partial^4}{\partial m^2 \partial u^2}.
\eeq 
Note that there are at least first order $u$-derivative in each terms. Therefore we can evaluate the higher order corrections to the quantum periods by using the $u$-derivative of $\Pi_{\gamma}^{(0)}$,
\beq
 \partial_{u}\Pi_{\alpha}^{(0)} = \frac{4\pi}{\Lambda_2}\lr{e^{-iq_2} - e^{-iq_1}}^{-\frac{1}{2}}\lr{e^{iq_2} - e^{iq_1}}^{-\frac{1}{2}}\ _2F_1\lrk{\frac{1}{2},\frac{1}{2},1,\frac{\lr{e^{iq_1} - e^{-iq_1}}\lr{e^{-iq_2} - e^{iq_2}}}{\lr{e^{-iq_2} - e^{-iq_1}}\lr{e^{iq_1} - e^{iq_2}}}},
\eeq
\beq
 \partial_{u}\Pi_{\beta}^{(0)} = \frac{4\pi}{\Lambda_2}\lr{e^{-iq_1} - e^{iq_1}}^{-\frac{1}{2}}\lr{e^{-iq_2} - e^{iq_2}}^{-\frac{1}{2}}\ _2F_1\lrk{\frac{1}{2},\frac{1}{2},1,\frac{\lr{e^{iq_2} - e^{iq_1}}\lr{e^{-iq_1} - e^{-iq_2}}}{\lr{e^{-iq_1} - e^{iq_1}}\lr{e^{iq_2} - e^{-iq_2}}}},
\eeq
where $q_1, q_2$ are the solutions to $u = V(q)$ (see Fig.\ref{fig:pot}).
In Table.\ref{tab:compare}, we compare the numerical results of $m_i^{(n)}$, calculated by the TBA equations, to $\Pi_{\gamma}^{(n)}$, calculated by the differential operators.

\begin{table}[tp]
\begin{center}
\hspace*{-0.4in} 
\begin{tabular}{ccccc} \hline
n & $(-1)^n\Pi_{\alpha}^{(n)}$ & $Z_1^{(n)}$, $Z_3^{(n)}$ & $i\Pi_{\beta}^{(n)}$ & $Z_2^{(n)}$, $Z_4^{(n)}$  \\ \hline
1 & $-\underline{0.502484}07$ & $-\underline{0.502484}11$ & $-\underline{0.239513}02$ & $-\underline{0.239513}16$ \\ 
2 & \hspace*{0.13in}$\underline{0.312000}29$ & \hspace*{0.13in}$\underline{0.312000}19$ & \hspace*{0.13in}$\underline{0.0158874}4$ & \hspace*{0.13in}$\underline{0.0158874}3$ \\ \hline
\end{tabular}
\caption{The numerical results of the coefficients of the quantum periods ($\Lambda = 4$, $u = 1$, $m = 1/8$). The numerical calculation in the TBA equations is done by Fourier discretization with $2^{18}$ points and a cutoff of the integrals $(-L, L)$ where $L = 40 + \log\lr{2\pi}$.}
\label{tab:compare}
\end{center}
\end{table}

\subsection{Effective central charge and one-loop beta function}
\label{sec:ceffb1}
By using the TBA equations, we can calculate the effective central charge $c_{\mathrm{eff}} = c - 24\Delta_{min}$ of the underlying 2d CFT , where $c$ is the central charge of the Virasoro algebra and $\Delta_{min}$ is the minimum eigenvalue of the Virasoro operator $L_0$. For (\ref{eq:TBA}), $c_{\mathrm{eff}}$ is given by \cite{ceff1, ceff2}
\begin{equation}
\label{eq:ceff}
  c_{\mathrm{eff}} = \frac{6}{\pi^2}\sum_{i = 1}^{4}Z_i\int_{\mathbb{R}}e^{\theta}L_i(\theta)d\theta = 4 + \frac{3}{\pi^2}\sum_{i=1}^{4}\left(\varepsilon_i^{\star}\log(1+e^{\varepsilon_i^{\star}}) + 2{\mathrm{Li}}_2(-e^{\varepsilon_i^{\star}})\right),
\end{equation}
where 
\beq
\varepsilon_i^{\star} = \lim_{\theta \rightarrow -\infty}\varepsilon_i(\theta).
\eeq
In $\theta \rightarrow -\infty$ limit (or equivalently, $\epsilon \rightarrow \infty$ limit), the TBA equations (\ref{eq:TBA}) lead
\beq
\begin{split}
 e^{-\varepsilon_1^{\star}} &= e^{-\varepsilon_3^{\star}} = \left(1 + e^{-\varepsilon_2^{\star}}\right)^{\frac{1}{2}}\left(1 + e^{-\varepsilon_4^{\star}}\right)^{\frac{1}{2}}, \\
 e^{-\varepsilon_2^{\star}} &= e^{-\varepsilon_4^{\star}} = \left(1 + e^{-\varepsilon_1^{\star}}\right)^{\frac{1}{2}}\left(1 + e^{-\varepsilon_3^{\star}}\right)^{\frac{1}{2}},
\end{split}
\eeq
and therefore
\beq
\label{eq:stars}
\begin{split}
 e^{-\varepsilon_1^{\star}} &= \left(1 + e^{-\varepsilon_2^{\star}}\right), \\
 e^{-\varepsilon_2^{\star}} &= \left(1 + e^{-\varepsilon_1^{\star}}\right).
\end{split}
\eeq
In fact, there are no mathematically rigorous solutions to (\ref{eq:stars}) \cite{1403.6137, 1403.7613, 2002.06829}. But we can formally consider that $\varepsilon_i^{\star} \rightarrow -\infty$ are the solutions. Then
\beq
 \varepsilon_i^{\star}\log(1+e^{\varepsilon_i^{\star}}) \rightarrow 0
\eeq
and
\beq
 {\mathrm{Li}}_2(-e^{\varepsilon_i^{\star}}) \rightarrow 0.
\eeq
Therefore  we obtain
\beq
 c_{\mathrm{eff}} = 4.
\eeq
This result agrees with the numerical calculation.
\par
We can also compute the effective central charge from the quantum periods. The large $\theta$ expansion of the TBA equations (\ref{eq:allorder}), (\ref{eq:Eper}) leads to a relational expression between $c_{\mathrm{eff}}$ and the quantum periods,
\beq
 c_{\mathrm{eff}} = -i\frac{6}{\pi} \left[\Pi_{\alpha}^{(0)}\Pi_{\beta}^{(1)} - \Pi_{\alpha}^{(1)}\Pi_{\beta}^{(0)}\right] + \frac{6}{\pi^2} (Z_3 - Z_1)\int_{\mathbb{R}}e^{\theta}L_3d\theta.
\eeq
$\Pi_{\alpha}^{(0)}\Pi_{\beta}^{(1)} - \Pi_{\alpha}^{(1)}\Pi_{\beta}^{(0)}$ can be expressed only $\Pi_{\gamma}^{(0)}$ by using the differential operator $\mathcal{O}_{1}$ (\ref{eq:FOOP}). After some transpositions, we finally get the following relation,
\beq
\label{eq:PNP}
\begin{split}
 \left[\Pi_{\alpha}^{(0)}\frac{\partial\Pi_{\beta}^{(0)}}{\partial u} - \frac{\partial \Pi_{\alpha}^{(0)}}{\partial u}\Pi_{\beta}^{(0)}\right] = &\ i\pi c_{\mathrm{eff}} - 3u \left[\Pi_{\alpha}^{(0)}\frac{\partial^2\Pi_{\beta}^{(0)}}{\partial u^2} - \frac{\partial^2 \Pi_{\alpha}^{(0)}}{\partial u^2}\Pi_{\beta}^{(0)}\right] \\ & - \frac{3}{2}m \left[\Pi_{\alpha}^{(0)}\frac{\partial^2\Pi_{\beta}^{(0)}}{\partial m \partial u} - \frac{\partial^2 \Pi_{\alpha}^{(0)}}{\partial m \partial u}\Pi_{\beta}^{(0)}\right] + i\frac{6}{\pi} (Z_3 - Z_1)\int_{\mathbb{R}}e^{\theta}L_3d\theta.
\end{split}
\eeq
In the massless case $m=0$, $\Pi_{\gamma}^{(0)}$ satisfies the second order Picard-Fuchs equation \cite{1705.09120} and the second, third and fourth terms in the r.h.s. of (\ref{eq:PNP}) become zero,
\beq
\left[\Pi_{\alpha}^{(0)}\frac{\partial\Pi_{\beta}^{(0)}}{\partial u} - \frac{\partial \Pi_{\alpha}^{(0)}}{\partial u}\Pi_{\beta}^{(0)}\right] = i\pi c_{\mathrm{eff}}.
\eeq
Because, at the massless point, the quantum periods agree with the ones for the pure SU(2) theory, the combination $\left[\Pi_{\alpha}^{(0)}\frac{\partial\Pi_{\beta}^{(0)}}{\partial u} - \frac{\partial \Pi_{\alpha}^{(0)}}{\partial u}\Pi_{\beta}^{(0)}\right]$ satisfies the Wronskian relation for the pure SU(2) theory \cite{hep-th/9506102}, 
\beq
\left[\Pi_{\alpha}^{(0)}\frac{\partial\Pi_{\beta}^{(0)}}{\partial u} - \frac{\partial \Pi_{\alpha}^{(0)}}{\partial u}\Pi_{\beta}^{(0)}\right] = 4\pi i.
\eeq
Therefore we obtain $c_{\mathrm{eff}} = 4$.
\par 
The relation (\ref{eq:PNP}) is also derived in the context of $\mathcal{N}=2$ gauge theories \cite{hep-th/9510129, hep-th/9510183}, while in these papers, the constant term in the r.h.s. is proportional to the one-loop beta functions of the $\mathcal{N}=2$ gauge theory. Therefore (\ref{eq:PNP}) indicates that $c_{\mathrm{eff}}$ for 2-dimensional CFT is proportional to the one-loop beta function for the SQCD. The similar relation also exists in the pure SU(2) case \cite{2002.06829}. 

\section{Summary and discussions}
\label{sec:CandO}
In this paper, we have investigated the exact WKB analysis for the quantum Seiberg-Witten curve of 4-dimensional $\mathcal{N} = 2\ SU(2)\ N_f = 2$ gauge theory with flavor symmetry and derived the TBA equations satisfied by the Borel resummations of the quantum periods in several parameter regions. We have also computed the effective central charge of the TBA equations and found the proportionality between the effective central charge and the one-loop beta function of the SQCD.
\par
As future works, we want to derive the TBA equations for other gauge theories whose quantum Seiberg-Witten curves form the \Schro type differential equations (e.g. $SU(2)$ with $N_f \leq 4$ \cite{1705.09120, 1604.05520}, $\mathcal{N}^{\ast} = 2$ theory \cite{1501.05671, 1108.0300, 1404.7378}). One of the possible way to derive the TBA equations is the ODE/IM correspondence proposed in \cite{hep-th/9812211}. For example, the TBA equations for pure $SU(2)$ case have already derived in \cite{1908.08030} by using the ODE/IM correspondence. Another possible way is using the integral equations proposed by Gaiotto, Moore and Neitzke in \cite{0807.4723}. The conformal limit of these integral equations becomes the TBA equations \cite{1403.6137} and the author argued that these TBA equations calculate the quantum periods. This argument was numerically demonstrated for pure $SU(2)$ case \cite{1908.07065} and showed in \cite{DGA1}. It is also interesting to study the relation to \cite{DGA2}, which studied the same Riemann-Hilbert problem we considered but used different methods. More ambitious generalization is the $\mathcal{N} = 2$ supersymmetric gauge theories with higher rank gauge group, whose quantum Seiberg-Witten curves form higher-order differential equations (for pure $SU(N)$ case \cite{0911.2396} and with matters \cite{1103.4843}). A good starting point is the $A_n$-type ODE studied in \cite{hep-th/0008039}, which relates to the quantum Seiberg-Witten curve for the $(A_n, A_m)$-type Argyres-Douglas theories \cite{1707.03596}. 
\par
It is also interesting to apply the TBA equations to study black hole physics. In \cite{2006.06111}, the authors claimed that the quasinormal mode frequencies for black holes are determined by the Bohr-Sommerfeld quantization condition for the quantum periods of 4-dimensional $\mathcal{N} = 2\ SU(2)$ gauge theory with $N_f = 2, 3$ and explicitly demonstrated at some lower levels. The more we included the higher order collections of the quantum periods, the more the spectrum obtained from the Bohr-Sommerfeld quantization condition matched to the true value. Therefore we expect that the Borel resummations of the quantum periods provide more precision.

\section*{Acknowledgements}
We would like to thank Katsushi Ito for valuable discussions and comments.


\begin{thebibliography}{50}
\bibitem{Pre}
N.~Seiberg,
Phys. Lett. B \textbf{206} (1988) 630-639 



\bibitem{hep-th/9407087}
N.~Seiberg and E.~Witten,
Nucl. Phys. B \textbf{426}, 19-52 (1994)
[arXiv:hep-th/9407087 [hep-th]].

\bibitem{hep-th/9408099}
N.~Seiberg and E.~Witten,
Nucl. Phys. B \textbf{431}, 484-550 (1994)
[arXiv:hep-th/9408099 [hep-th]].

\bibitem{hep-th/9602082}
F.~Ferrari and A.~Bilal,
Nucl. Phys. B \textbf{469}, 387-402 (1996)
[arXiv:hep-th/9602082 [hep-th]].

\bibitem{hep-th/9605101}
A.~Bilal and F.~Ferrari,
Nucl. Phys. B \textbf{480}, 589-622 (1996)
[arXiv:hep-th/9605101 [hep-th]].

\bibitem{hep-th/9611012}
F.~Ferrari,
Nucl. Phys. B Proc. Suppl. \textbf{55}, no.2, 245-252 (1997)
[arXiv:hep-th/9611012 [hep-th]].

\bibitem{hep-th/9706145}
A.~Bilal and F.~Ferrari,
Nucl. Phys. B \textbf{516}, 175-228 (1998)
[arXiv:hep-th/9706145 [hep-th]].

\bibitem{hep-th/0206161}
N.~A.~Nekrasov,
Adv. Theor. Math. Phys. \textbf{7}, no.5, 831-864 (2003)
[arXiv:hep-th/0206161 [hep-th]].

\bibitem{hep-th/0306238}
N.~Nekrasov and A.~Okounkov,
Prog. Math. \textbf{244}, 525-596 (2006)
[arXiv:hep-th/0306238 [hep-th]].

\bibitem{0910.5670}
A.~Mironov and A.~Morozov,
JHEP \textbf{04}, 040 (2010)
[arXiv:0910.5670 [hep-th]].

\bibitem{0908.4052}
N.~A.~Nekrasov and S.~L.~Shatashvili,
[arXiv:0908.4052 [hep-th]].

\bibitem{0911.2396}
A.~Mironov and A.~Morozov,
J. Phys. A \textbf{43}, 195401 (2010)
[arXiv:0911.2396 [hep-th]].

\bibitem{1103.4843}
Y.~Zenkevich,
Phys. Lett. B \textbf{701}, 630-639 (2011)
[arXiv:1103.4843 [math-ph]].

\bibitem{1705.09120}
K.~Ito, S.~Kanno and T.~Okubo,
JHEP \textbf{08}, 065 (2017)
[arXiv:1705.09120 [hep-th]].

\bibitem{1804.04815}
K.~Ito and T.~Okubo,
Nucl. Phys. B \textbf{934}, 356-379 (2018)
[arXiv:1804.04815 [hep-th]].

\bibitem{1903.00168}
K.~Ito, S.~Koizumi and T.~Okubo,
Phys. Lett. B \textbf{792}, 29-34 (2019)
[arXiv:1903.00168 [hep-th]].

\bibitem{2001.08891}
K.~Ito, S.~Koizumi and T.~Okubo,
Nucl. Phys. B \textbf{954}, 115004 (2020)
[arXiv:2001.08891 [hep-th]].

\bibitem{0910.4431}
H.~Awata and Y.~Yamada,
JHEP \textbf{01}, 125 (2010)
[arXiv:0910.4431 [hep-th]].

\bibitem{1008.0574}
H.~Awata, H.~Fuji, H.~Kanno, M.~Manabe and Y.~Yamada,
Adv. Theor. Math. Phys. \textbf{16}, no.3, 725-804 (2012)
[arXiv:1008.0574 [hep-th]].

\bibitem{1504.08324}
A.~K.~Kashani-Poor and J.~Troost,
JHEP \textbf{08}, 160 (2015)
[arXiv:1504.08324 [hep-th]].

\bibitem{1604.05520}
S.~K.~Ashok, D.~P.~Jatkar, R.~R.~John, M.~Raman and J.~Troost,
JHEP \textbf{07}, 115 (2016)
[arXiv:1604.05520 [hep-th]].

\bibitem{1811.04812}
K.~Ito, M.~Mari\~no and H.~Shu,
JHEP \textbf{01}, 228 (2019)
[arXiv:1811.04812 [hep-th]].

\bibitem{1803.02320}
A.~Grassi and J.~Gu,
JHEP \textbf{02}, 060 (2019)
[arXiv:1803.02320 [hep-th]].

\bibitem{1806.01407}
A.~Grassi and M.~Mari\~no,
SIGMA \textbf{15}, 025 (2019)
[arXiv:1806.01407 [hep-th]].



\bibitem{Vor}
A.~Voros,
Ann. I.H.P. 39 (1983) 211 



\bibitem{1908.07065}
A.~Grassi, J.~Gu and M.~Mari\~no,
JHEP \textbf{07}, 106 (2020)
[arXiv:1908.07065 [hep-th]].

\bibitem{2002.06829}
K.~Imaizumi,
Phys. Lett. B \textbf{806}, 135500 (2020)
[arXiv:2002.06829 [hep-th]].

\bibitem{1910.09406}
K.~Ito and H.~Shu,
J. Phys. A \textbf{53}, 335201 (2020) 33
[arXiv:1910.09406 [hep-th]]

\bibitem{1906.04271}
L.~Hollands and A.~Neitzke,
Commun. Math. Phys. \textbf{380}, no.1, 131-186 (2020)
[arXiv:1906.04271 [hep-th]].

\bibitem{2012.15658}
F.~Yan,
[arXiv:2012.15658 [hep-th]].

\bibitem{0907.3987}
D.~Gaiotto, G.~W.~Moore and A.~Neitzke,
[arXiv:0907.3987 [hep-th]].



\bibitem{DP}
E.~Delabaere and F.~Pham,
Annales de l’IHP {\bf 71} (1999) 1-94

\bibitem{cr1}
K.~Iwaki and T.~Nakanishi,
J. Phys. A {\bf 47}, 474009 (2014)
[arXiv:1401.7094 [math.CA]]



\bibitem{1002.2459}
L.~F.~Alday, J.~Maldacena, A.~Sever and P.~Vieira,
J. Phys. A \textbf{43}, 485401 (2010)
[arXiv:1002.2459 [hep-th]].

\bibitem{0807.4723}
D.~Gaiotto, G.~W.~Moore and A.~Neitzke,
Commun. Math. Phys. \textbf{299}, 163-224 (2010)
[arXiv:0807.4723 [hep-th]].

\bibitem{hep-th/9511154}
P.~C.~Argyres, M.~R.~Plesser, N.~Seiberg and E.~Witten,
Nucl. Phys. B \textbf{461}, 71-84 (1996)
[arXiv:hep-th/9511154 [hep-th]].


\bibitem{ceff1}
A.~B.~Zamolodchikov, 
Nucl. Phys. B \textbf{342}, 695-720 (1990).

\bibitem{ceff2}
T.~R.~Klassen and E.~Melzer, 
Nucl. Phys. B \textbf{338} 485-528 (1990).


\bibitem{1403.6137}
D.~Gaiotto,
[arXiv:1403.6137 [hep-th]].


\bibitem{hep-th/9506102}
M.~Matone,
Phys. Lett. B \textbf{357} 342-348 (1995)
[arXiv:hep-th/9506102 [hep-th]].



\bibitem{1403.7613}
S.~Cecotti and M.~Del Zotto,
J. Phys. A \textbf{47}, no.47, 474001 (2014)
[arXiv:1403.7613 [hep-th]].

\bibitem{hep-th/9510129}
J.~Sonnenschein, S.~Theisen and S.~Yankielowicz,
Phys. Lett. B \textbf{367}, 145-150 (1996)
[arXiv:hep-th/9510129 [hep-th]].

\bibitem{hep-th/9510183}
T.~Eguchi and S.~K.~Yang,
Mod. Phys. Lett. A \textbf{11}, 131-138 (1996)
[arXiv:hep-th/9510183 [hep-th]].

\bibitem{1501.05671}
G.~Ba\c{s}ar and G.~V.~Dunne,
JHEP \textbf{02}, 160 (2015)
[arXiv:1501.05671 [hep-th]].

\bibitem{1108.0300}
W.~He,
J. Math. Phys. \textbf{56}, no.7, 072302 (2015)
[arXiv:1108.0300 [math-ph]].

\bibitem{1404.7378}
A.~K.~Kashani-Poor and J.~Troost,
JHEP \textbf{08}, 117 (2014)
[arXiv:1404.7378 [hep-th]].

\bibitem{hep-th/9812211}
P.~Dorey and R.~Tateo,
J. Phys. A \textbf{32}, L419-L425 (1999)
[arXiv:hep-th/9812211 [hep-th]].

\bibitem{1908.08030}
D.~Fioravanti and D.~Gregori,
Phys. Lett. B \textbf{804}, 135376 (2020)
[arXiv:1908.08030 [hep-th]].


\bibitem{DGA1}
Dylan~G.L.~Allegretti,
J. Topol. \textbf{12}, 1031-1068 (2019)
[arXiv:1802.05479 [math.CA]]

\bibitem{DGA2}
Dylan~G.L.~Allegretti,
Adv. Math. \textbf{380} (2021)
[arXiv:1912.05938 [math.AG]]


\bibitem{hep-th/0008039}
P.~Dorey, C.~Dunning and R.~Tateo,
J. Phys. A \textbf{33}, 8427-8442 (2000)
[arXiv:hep-th/0008039 [hep-th]].

\bibitem{1707.03596}
K.~Ito and H.~Shu,
JHEP \textbf{08}, 071 (2017)
[arXiv:1707.03596 [hep-th]].

\bibitem{2006.06111}
G.~Aminov, A.~Grassi and Y.~Hatsuda,
[arXiv:2006.06111 [hep-th]].
\end{thebibliography}
\end{document}